\documentclass[sigconf, 9pt]{acmart}
\usepackage[linesnumbered,ruled,vlined]{algorithm2e}
\usepackage{enumitem}
\usepackage{algpseudocode}
\usepackage{amsmath}
\usepackage{mathtools, nccmath}
\usepackage{eucal}
\usepackage{bm}

\usepackage{titlesec}
\titlespacing*{\section}{0pt}{0.1\baselineskip}{0.05\baselineskip}
\titlespacing*{\subsection}{0pt}{0.1\baselineskip}{0.05\baselineskip}
\titlespacing*{\subsubsection}{0pt}{0.1\baselineskip}{0.05\baselineskip}

\newtheoremstyle{sig}
  {}
  {}
  {\itshape}
  {}
  {\scshape}
  {.}
  {.5em}
  {#1 #2\thmnote{\quad(#3)}}

\theoremstyle{sig}

\AtBeginDocument{%
  \providecommand\BibTeX{{%
    \normalfont B\kern-0.5em{\scshape i\kern-0.25em b}\kern-0.8em\TeX}}}

\setcopyright{acmcopyright}
\copyrightyear{2024}
\acmYear{2024}

\acmConference[DAC '24]{the 61st
ACM/IEEE Design Automation Conference}{June 23-27,
  2024}{San Francisco, CA, USA}
\begin{document}

\title{Go Beyond Black-box Policies: Rethinking the Design of Learning
Agent for Interpretable and Verifiable HVAC Control}

\author{Zhiyu An}
\affiliation{%
  \institution{University of California, Merced}
  \country{}
}
\email{zan7@ucmerced.edu}

\author{Xianzhong Ding}
\affiliation{%
  \institution{University of California, Merced}
  \country{}
}
\email{xding5@ucmerced.edu}

\author{Wan Du}
\affiliation{%
  \institution{University of California, Merced}
  \country{}
}
\email{wdu3@ucmerced.edu}


\renewcommand{\shortauthors}{An et al.}

\begin{abstract}
Recent research has shown the potential of Model-based Reinforcement Learning (MBRL) to enhance energy efficiency of Heating, Ventilation, and Air Conditioning (HVAC) systems. However, existing methods rely on black-box thermal dynamics models and stochastic optimizers, lacking reliability guarantees and posing risks to occupant health.
In this work, we overcome the reliability bottleneck by redesigning HVAC controllers using decision trees extracted from existing thermal dynamics models and historical data. Our decision tree-based policies are deterministic, verifiable, interpretable, and more energy-efficient than current MBRL methods. First, we introduce a novel verification criterion for RL agents in HVAC control based on domain knowledge. Second, we develop a policy extraction procedure that produces a verifiable decision tree policy. We found that the high dimensionality of the thermal dynamics model input hinders the efficiency of policy extraction. To tackle the dimensionality challenge, we leverage importance sampling conditioned on historical data distributions, significantly improving policy extraction efficiency. Lastly, we present an offline verification algorithm that guarantees the reliability of a control policy.
Extensive experiments show that our method saves 68.4\% more energy and increases human comfort gain by 14.8\% compared to the state-of-the-art method, in addition to an 1127$\times$ reduction in computation overhead.
Our code and data are available at \color{blue}https://github.com/ryeii/Veri\_HVAC\color{black}.

\end{abstract}






\maketitle
\vspace{-10pt}
\section{Introduction}
Efficient control of Heating, Ventilation, and Air Conditioning (HVAC) systems stands as a critical cornerstone of building operations, exerting a direct influence on energy consumption and the comfort of occupants \cite{building_energy}. Model-free Reinforcement Learning (MFRL) has been extensively explored for HVAC control \cite{wei2017deep, ding2019octopus}, demonstrating promising performance. Nevertheless, the inherent data-hungry nature of MFRL, reliant on trial-and-error interactions with real-world buildings to learn optimal HVAC control policies, presents a significant obstacle to practical deployment.

Recent research endeavors have shown the potential of Model-based Reinforcement Learning (MBRL)~\cite{zhang2019building, ding2020mb2c, an2023clue} for HVAC control, offering high data efficiency. However, widespread integration of MBRL in the HVAC industry is hampered by concerns about its reliability and interoperability~\cite{paleyes2022challenges, an2024reward}. Its black-box thermal dynamics models and stochastic optimization pose significant barriers to safety and understanding behavior in practice~\cite{li2023trustworthy, brunke2022safe}.

In response to the above challenge, this paper addresses the reliability bottleneck associated with MBRL in HVAC control. We redesign HVAC controllers by employing policy-extracted decision trees. The output is a set of policies that are not only verifiable and interpretable \cite{bastani2018verifiable}, but also surpass current MBRL methods in terms of performance. This paper introduces a novel and domain-specific verification criterion for HVAC controllers. We also present a policy extraction procedure that yields decision tree policies, which are not only interpretable but also suitable for direct deployment in real-world building environments. To cope with the high-dimensional nature of HVAC control problems, we leverage importance sampling conditioned on historical data distributions, a technique that significantly enhances the efficiency of policy extraction. To specifically evaluate the reliability of HVAC control policies, we introduce an offline verification algorithm that employs decision queries and probabilistic verification techniques to assess the safety and performance of the policies under various conditions.

We conduct extensive experiments to validate the effectiveness of our approach in overcoming the reliability challenge in HVAC control. Our contributions aim to pave the way for the practical deployment of learning-based HVAC controllers, offering not only superior performance but also the crucial assurance of safety and interpretability in critical building infrastructure systems. In summary, this paper makes the following significant contributions:
\begin{itemize}[topsep=0pt,leftmargin=*]
\item We address the critical reliability challenge in HVAC control by introducing a novel and domain-specific verification criterion for RL agents, ensuring safety and dependability in their operation.

\item We develop a policy extraction procedure that produces interpretable decision tree policies, enabling straightforward deployment in real-world building environments while significantly outperforming current MBRL methods.

\item We present an innovative offline verification algorithm that evaluates the reliability of HVAC control policies through decision queries and probabilistic verification, offering a robust approach to ensure the safety of RL agents in HVAC systems.

\item Extensive experiments show the efficacy of proposed methods.
\end{itemize}

\section{Preliminary and Motivation}

\subsection{MBRL for HVAC Control}

We formulate HVAC control problem as a discrete-time Markov Decision Process (MDP) $\mathcal{M}:\{\mathcal{S}, \mathcal{D}, \mathcal{A}, r, f, \gamma\}$, consisting of the state space $\mathcal{S}$, the disturbance space $\mathcal{D}$, the action space $\mathcal{A}$, the reward function $r:\mathcal{S}\times\mathcal{A}\rightarrow\mathbb{R}$, the dynamics function $f(s'|s, d, a)$ and discount factor $\gamma$. At each time step $t$, the system is in state $s_t\in\mathcal{S}$, subject to disturbances $d_t\in\mathcal{D}$, executes some action $a_t\in\mathcal{A}$, receives reward $r_t=r(s_t, a_t)$, and transitions to the next state $s_{t+1}$ according to the dynamics function $s_{t+1}\sim f(s_t, d_t, a_t)$. At each time step, the control agent applies a policy $\pi:(\mathcal{S}\times\mathcal{D})\rightarrow\mathcal{A}$ to choose the action that maximizes the discounted sum of future rewards, given by $\sum_{t=0}^\infty\gamma^tr(s_t, a_t)$, where $\gamma\in[0, 1]$ is the discount factor that prioritizes near-term rewards.

MBRL-based control approximates the optimal policy using two components: the dynamics model and the controller. 
The dynamics model is a regression model that learns the discrete-time thermal dynamics of the building system by training on a set of historical data $\{(s_t, d_t, a_t, s_{t+1})\}_N$. It predicts $s_{t+1}$ based on $(s_t, d_t, a_t)$, and the predictions are then used by the controller to choose the optimal action. 
The controller solves the following optimization problem using a stochastic optimizer such as Random Shooting (RS)~\cite{zhang2019building} and Model Predictive Path Integral (MPPI)~\cite{an2023clue} algorithm:
\begin{equation}\label{Eq:background optimization problem}
    a[:] = \arg\max_{a[:]}\sum_{t=1}^H\gamma^tr(\hat{f}(s_t, d_t, a_t), a_{t-1})
\end{equation}
where $\hat{f}$ is the learned dynamics model. The controller picks the action sequence of size $H$ that maximizes the cumulative discounted rewards of the future $H$ time steps. In practice, the controller executes only the first action from the sequence and then solves Eq. \ref{Eq:background optimization problem} again in the next time step with the updated state information. 

\textbf{States.} The zone state variable is the temperature of the controlled thermal zone, which depends on our control action and is used to calculate the building system reward. It is used as part of the input and the only output of the system dynamics model.

\textbf{Disturbances.} The disturbances comprise variables that do not depend on the control action of the HVAC system, including weather conditions and occupancy. The variables of the state and the disturbances are specified in Table \ref{table_variables}.

\textbf{Actions.} The action is the temperature setpoint of the controlled thermal zone. Each zone is associated with a heating setpoint and a cooling setpoint, resulting in an action dimension of $2$. In our experimental platform, the setpoint for the HVAC system is an integer in $[15^\circ C, 23^\circ C]$ for heating, and $[21^\circ C, 30^\circ C]$ for cooling.

\textbf{Rewards.}\label{Sec: reward design}
We adopt the reward function described in \cite{jimenez2021sinergym}, represented by Eq. \ref{Eq: reward}. The comfort zone is defined as $[\underbar{z}, \overline{z}]$, which represents the bounds for the zone temperature. At each time step $t$, $E_t$ represents the total energy consumption, which is estimated by taking the L1 norm of the difference between the setpoint and the setpoint corresponding to the HVAC being turned off~\cite{chen2019gnu}. To balance the relative importance of comfort and energy consumption, we used a weight variable $w_e\in[0, 1]$. 

\begin{equation} \label{Eq: reward}
    r(s_t) = - w_eE_t - (1-w_e)(|s_t - \overline{z}|_++|s_t-\underbar{z}|_+)
\end{equation}

We set $w_e = 1e-2$ during occupied periods and $w_e = 1$ during unoccupied periods. Comfort zones are $[20^\circ C, 23.5^\circ C]$ for the winter and $[23^\circ C, 26^\circ C]$ for the summer.

\begin{table}[]
    \begin{tabular}{r|l}
    \hline
       State $s_t$ & Zone Air Temperature ($^\circ C$) \\
    \hline
       Disturbances $d_t$ & Outdoor Air Drybulb Temperature ($^\circ C$) \\
       & Outdoor Air Relative Humidity (\%) \\
       & Site Wind Speed ($m/s$) \\
       & Site Total Radiation Rate Per Area ($W/m^2$) \\
       & Zone People Occupant Count ($No.$) \\
    \hline
    \end{tabular}
    \caption{State and disturbance variables.}
    \label{table_variables}
\end{table}

\subsection{Motivation Experiments}
\begin{figure}
\includegraphics[width=\columnwidth]{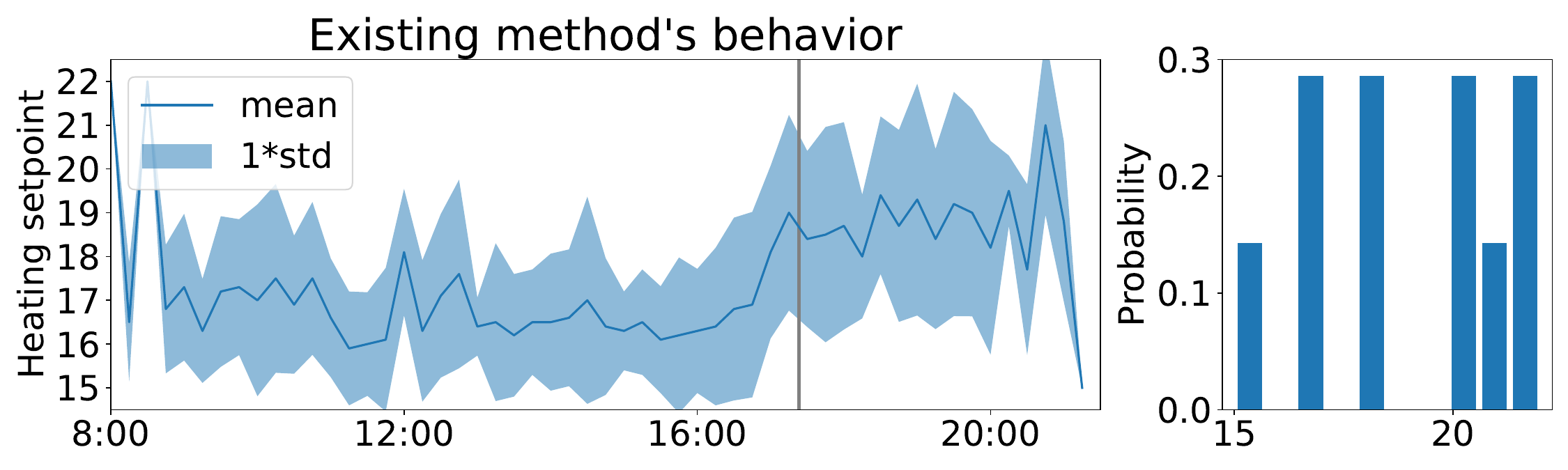}
\caption{Left: the distribution of setpoints over 10 runs on a fixed set of disturbances of one day. Right: the distribution of setpoints in the left figure.}
\label{fig: Jan1_pittsburgh}
\Description[]{}
\end{figure}

To understand the decision uncertainty of the state-of-the-art MBRL method \cite{ding2020mb2c}, we perform simulations using EnergyPlus for a building with five zones, detailed in Section \ref{sec: Platform and Implementation Details}. We implement the methodology described in \cite{ding2020mb2c} as a conventional MBRL approach.

\textbf{Experiment results.}
Fig. \ref{fig: Jan1_pittsburgh} shows the heating setpoint behavior of of existing MBRL method \cite{ding2020mb2c} over one day. 
The left subfigure shows the mean heating setpoint values as a function of time, from 8:00 to over 22:00. The mean setpoint fluctuates throughout the day within a range of 15°C to 22°C. The shading around the mean represents one standard deviation, indicating the variability of the setpoint selection. We run the experiments 10 times during a simulated day while maintaining fixed disturbances. Over the 10 runs, we calculated the deviation of each time point. The spread of the shaded area suggests that there is considerable variability in the heating setpoint, implying that the method does not consistently choose the same setpoint, even under the same weather scenario. 

The right bar chart displays the probability distribution of the heating setpoint choices of the existing MBRL method \cite{ding2020mb2c}. There are six bars corresponding to the same time shown in the black line in the left figure, with the heights representing the probability of the setpoint being selected. The distribution is relatively even, with no single setpoint having a dominant probability, which illustrates the stochastic nature of the existing MBRL method's behavior. 

\textbf{Challenge: Stochasticity of Black-box MBRL Policy.}
When considering the same weather scenario, it becomes evident that the setpoint decisions made by the existing black-box MBRL policy exhibit significant stochasticity. The distribution of the existing method's setpoints in a one-time step revealed that it has >$10\%$ probability of choosing both the highest setpoint ($22^\circ C$) and the lowest setpoint ($15^\circ C$). If the true optimal range of setpoints spans the entire setpoint spectrum, the only plausible explanation would be that the choice of setpoint does not have a discernible impact. However, this cannot hold true since different setpoints lead to varying energy consumption. Consequently, the inherent stochasticity of the existing method renders it inherently suboptimal.

Based on the above observations, our primary goal is to address the limitations associated with the policy stochasticity in existing MBRL methods. We aim to develop a novel approach characterized by determinism, where every input corresponds to a decision with a certainty of 100\%. This eliminates safety concerns arising from the unpredictability of controllers. Furthermore, it enables us to determine and evaluate its behavior in unseen environments before deployment, i.e., verification of the underlying policy. This is critical for identifying and mitigating potential controller faults, ensuring consistent and accurate operation.

\section{Proposed Approach}


\begin{figure*}
\includegraphics[width=\linewidth]{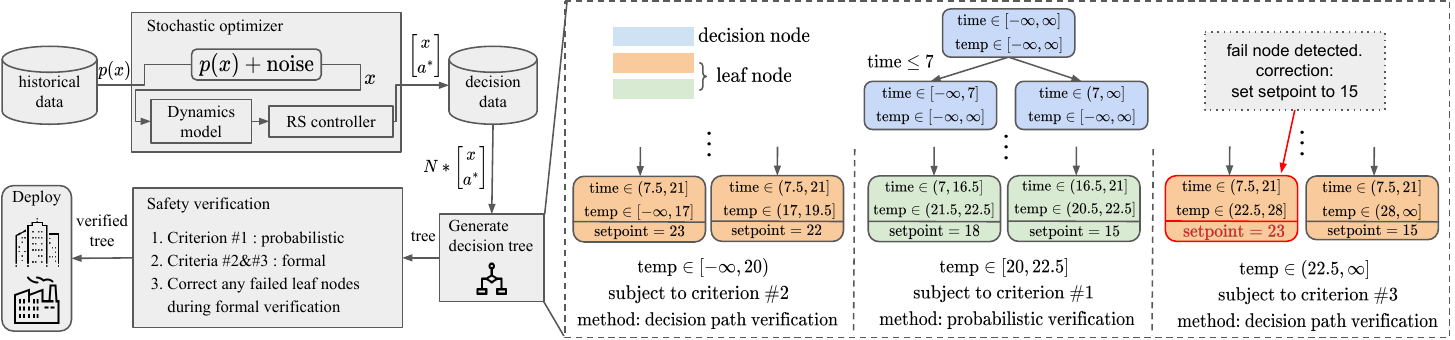}
\caption{Left: our proposed procedure. Right: an illustration of a DT with two variables (time and temp). The leaf nodes are classified into three categories based on temperature. The decision path verification algorithm detects and corrects failed nodes.}
\label{fig: overview}
\Description[]{}
\end{figure*}

Our proposed procedure is illustrated on the left side of Fig.\ref{fig: overview}. The procedure starts from the historical data of the building thermal dynamics, extracts a decision tree policy, verifies the safety of the policy, and deploys it to the building edge device.
The rest of this section describes our approach in three parts.
First, we introduce verification criteria for HVAC control policies based on the domain knowledge about HVAC operation safety. 
Then, we describe our policy extraction procedure that automatically constructs the decision tree policy using the learned black-box system dynamics model. 
Lastly, we describe two algorithms to formally and probabilistically verify the extracted policy against the verification criteria.

\subsection{Verification Criteria For HVAC Control} \label{sec: Verification Criteria For HVAC Control}

We focus on the precise air temperature control of a thermal zone during occupied hours. 
Hence, we define the set of "safe" states as $s\in [\underbar{z}, \overline{z}]$, where $[\underbar{z}, \overline{z}]$ is the predefined comfort range. Given the said safety criterion, we aim to construct a set of verification criteria to check the policy's output (the setpoint) for an infinite set of inputs (the building state and disturbances). Ideally, these criteria should be the tightest boundaries on the policy outputs. Satisfying the boundaries provides a safety guarantee, while not interfering with the effective operation of the data-driven policy. 

To construct these criteria, we first divide the input building states into three subsets by domain knowledge, then develop verification for each of the input subsets. When the zone temperature violates the comfort range, it is always desirable for the HVAC system to provide responsive heating/cooling in an effort to correct the temperature. The amount of heating/cooling it should provide is, however, hard to determine. For instance, if the disturbances rapidly cool the zone, and the zone temperature is only $0.5^\circ C$ too warm than the comfort range, then blindly setting to the lowest setpoint for the next time step (15 minutes) without considering the changing rate of zone temperature can result in the zone temperature dropping below the comfort range. To mitigate the risk of under/overshooting, we only bound the setpoint to be above the zone temperature when the zone is too cold, and below the zone temperature when the zone is too warm while allowing the MBRL agent to determine the exact setpoint in that range.

When the zone temperature is within the comfort range, we only need to make sure that the setpoint selected by the policy keeps the zone temperature within the comfort range in the future. This is a sequential decision problem that is difficult to manually solve. It is the reason that MBRL is applied in the first place. In addition, the stochasticity of the disturbances makes it difficult to verify all possible combinations of disturbances. Hence, we adopt probabilistic verification~\cite{bacci2022verified, bacci2022formal}, which estimates the probability of the system reaching the fail states within $H$ time steps into the future. Formally, we construct a forward reachability tube~\cite{landers2023deep} in Eq.\ref{eq: forward reachable tube 1}, i.e., all the possible states reachable within $H$ time steps given the policy $\pi$. We then estimate the safe probability and compare it with the probability threshold $l$ specified by the building manager. Finally, we combine all three criteria and define them in Eq.~\ref{eq: verified suffice}.

Note that verification criteria \#2 and \#3 are stronger than \#1 because they are not probabilistic - satisfying these criteria provides a $100\%$ guarantee on the policy behavior. Thus, our three-component criteria are stronger than applying probabilistic verification to the entire input space. 
\vspace{-5pt}
\begin{multline}\label{eq: forward reachable tube 1}
    \mathcal{R}^+(s_0)|_\pi^H = \bigcup_{t = 0}^{H} \; \{ s_t \in \mathcal{S} \;|\; s_{t+1} = f(s_t,d_t,a_t),\\ a_t \sim \pi(s_t, d_t) \text{ for } t \in [0, H] \}
\end{multline}
\vspace{-10pt}
\begin{equation}\label{eq: verified suffice}
\text{verified}(\pi)\iff
    \begin{cases}
        \#1:~\mathbb{E}[\overline{z} \ge s_t \ge \underbar{z}]>l & \forall s_t \in \mathcal{R}^+(s_0)|_\pi^\infty,\\
        \#2:~\pi(s_t, d_t) < s_t & \text{if }s_t > \overline{z},\\
        \#3:~\pi(s_t, d_t) > s_t & \text{if }s_t < \underbar{z}\\
    \end{cases}
\end{equation}

\subsection{Policy Extraction using Black-box System Dynamics Model}

We start from the standard MBRL components~\cite{ding2020mb2c, an2023clue}: historical dataset $\mathcal{T}:\{(s, d, a, s')\}$ extracted from the building management systems (BMS), a system dynamics model $\hat{f}$ learned from the said dataset, and a stochastic optimizer $RS$.
Our goal is to produce a decision tree $T:(\mathcal{S}\times\mathcal{D})\rightarrow\mathcal{A}$ which takes the current zone temperature, current disturbances, and outputs a setpoint which will be actuated in the next time step. 
To do that, we employ a two-stage process. First, we construct a decision dataset $\Pi:\{(s, d, a)\}$ consisting of the policy input and the approximated optimal action. Then, we utilize the CART (Classification and Regression Trees) algorithm\cite{loh2011classification} to automatically construct a decision tree to fit the decision dataset. 
We now describe both procedures.

\subsubsection{Decision Dataset Generation} \label{Sec: Decision Dataset Generation}

The entries of $\Pi$ are produced by distilling the stochastic decisions of an MBRL HVAC controller into deterministic decisions. Given the system dynamics model, the MBRL controller uses a stochastic optimizer, e.g. random shooting, to approximate the optimal setpoint \cite{ding2020mb2c, an2023clue}. Let $\hat{a} = \hat{\pi}(s, d)$ be the setpoint approximated by the stochastic optimizer, $p(\hat{a})$ be its distribution obtained by Monte Carlo method. We define $a^*$ as the most frequent $a$ in $p(\hat{a})$ and append $(s, d, a^*)$ to the decision dataset.

Ideally, dataset $\Pi$ contains the optimal decision for all possible combinations of the inputs, which allows the learned policy to generalize to unseen states. However, this introduces a unique challenge. Empirically, the overhead to sample an optimal decision for one input averages to $500$ milliseconds with Intel i9-11900KF and GeForce RTX 3080Ti. If we tentatively divide each continuous input variable into $20$ bins (which is sparse, considering the outdoor temperature generally spans $0^\circ C - 36^\circ C$) and measure the resulting density using this sampling strategy, it will take $20^5$ samples to obtain $1$ sample per bin on average, which takes around $444$ hours.

\begin{figure}
\includegraphics[width=\columnwidth]{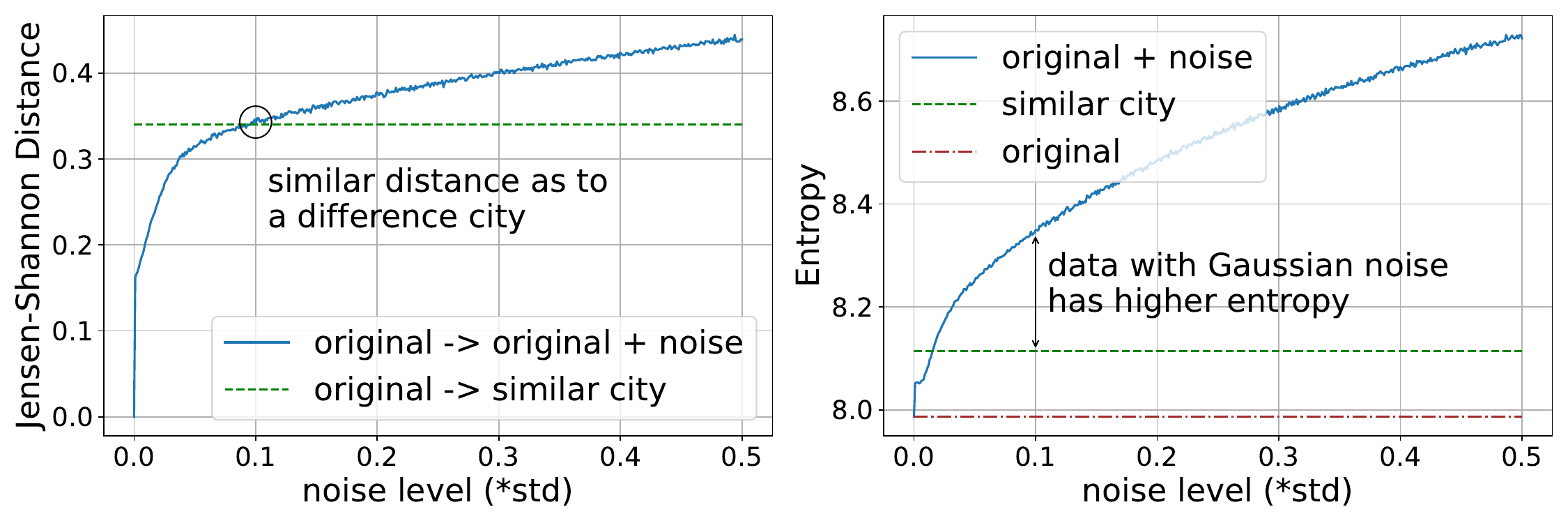}
\caption{Preliminary experiment to determine the appropriate noise level.}
\label{fig: JSD_vs_noise_level}
\Description[]{}
\end{figure}

Fortunately, for HVAC control, we do not need to sample every possible input. Recent work by An et al.~\cite{an2023clue} found that each city has a unique distribution of input states resulting from their unique weather profiles. In other words, some scenarios occur more frequently than others. Thus, sampling the optimal action of the more frequent scenarios provides more gain than sampling less frequent scenarios. One possible approach is to first calculate the bin-wise density of the historical data, but that costs $O(n^5)$ space complexity, where $n$ is the number of bins for each continuous variable in the input. Instead, we directly sample the historical data and add an element-wise Gaussian noise to each sample, formally described in Eq.\ref{Eq. noise}, where $X$ denotes the historical data.
\begin{equation}\label{Eq. noise}
    \widehat{p(x)} = X+\mathcal{N}\left(0, \text{noise\_level}\times\sqrt{\frac{\sum(x_i-\overline{x})^2}{|X|}}\right)
\end{equation}
Such data augmentation has to balance between two competing objectives. First, the noise has to be sufficiently large to allow the resulting policy to generalize to new, unseen inputs. Second, the noise should not be too large which causes the sampled distribution to lose resemblance to the original distribution, which decreases sample efficiency. To determine the appropriate noise level, we conducted a preliminary experiment in Pittsburgh and New York, both classified as climate category 4A by ASHRAE~\cite{standard2020ansi}, ensuring a fair comparison. We tested different noise levels from $0.01$ to $0.5$ and compared the Information Entropy and Jensen-Shannon Distance (JSD)\cite{nielsen2019jensen} of the original historical data distribution, distribution after adding noise, and the distribution of the other city. The result is shown in Fig.~\ref{fig: JSD_vs_noise_level}. The ideal noise level should result in a JSD lower than the other city and an entropy as large as possible. Based on the experiment result, we set the noise level to $[0.01, 0.09]$. After obtaining $\Pi$ using Monte Carlo method on $\widehat{p(x)}$, we proceed to construct the decision tree.

\subsubsection{Constructing the Decision Tree.}

The decision tree policy is an unweighted directed acyclic graph consisting of decision nodes and leaf nodes, as illustrated in Fig.\ref{fig: overview}. Each decision node is connected with two child nodes and contains a threshold value, which is compared with one element in the input vector. Then, the decision node calls either of its two child nodes based on the comparison result ($\le$ or $>$). Each leaf node contains a setpoint decision, which will be returned once called. The tree is constructed from $\Pi$ using a two-step process. First, we concatenate the elements of the input tuple $(s, d)$ of the decision dataset $\Pi$ to form a single input vector $x$, each element still represents its original meaning according to their relative index in the input vector before concatenation. Then, we fit a classification decision tree $T:\mathcal{X}\rightarrow\mathcal{A}$ which takes an input vector, computes the forward propagation, finds the resulting leaf node, and outputs the setpoint decision of that leaf node. Since each decision node only compares with one element in the input vector, the tree is fully interpretable and knowledgeable to human experts. Now, we can utilize the interpretable property of this decision tree policy to verify it in the next section.

\subsection{Verifying the Decision Tree Policy}

The verification criteria in Section~\ref{sec: Verification Criteria For HVAC Control} involves two types of verification. For criteria \#2 and \#3, we design a formal verification algorithm, verify the decision tree, and correct the failed cases by directly editing the decision tree. For criteria \#1, we apply probabilistic verification using Monte Carlo method.

\subsubsection{Verifying Criteria \#2 and \#3.}

When the zone temperature already violates the comfort range, we expect the controller to always make an effort to correct the zone temperature. 
Our goal is to provide a $100\%$ guarantee on the underlying policy behavior.
To do that, we develop a verification algorithm, shown in Alg.\ref{alg: Tree traverse verification}. The key idea is that each leaf node has a unique path from the root node. Since $T$ is surjective from $\mathcal{X}$ to $\mathcal{A}$, each leaf node must deterministically handle a subset of the input space. Thanks to the interpretability of decision tree, we can compute the said subset for each leaf node, identify the leaf nodes that handle the inputs of interest, and verify the setpoints of these leaf nodes. In words, Algorithm \ref{alg: Tree traverse verification} does the following: 1) it iterates through all leaf node, and compute the unique path that connects the root node to each leaf node; 2) for each path, it computes the union of the "boxes" on the values of the input vectors handled by the decision nodes along the path; 3) based on the box of the leaf node, it determines if this node will be called if the input belongs to the set in interest. Finally, we check if the decision of the leaf node complies with the criteria.
\begin{algorithm}
\caption{Decision path verification}
\label{alg: Tree traverse verification}
\For{each leaf node $T_i$}{
        $P = \{T_0, \cdots, T_i\}\gets$ extract path from $T_0$ to $T_i$\\
        $C = \mathbb{R}^{|\mathcal{X}|}\;\;$ \Comment{initialize state box boundaries}\\
        \For{$T_j \in (P\setminus T_i)$}{
            $C \gets C \bigcap \;(\text{input box of the rules from }T_j\text{ to }T_{j+1})$
        }
        \If{$C\subseteq ((s_t > \overline{z})\lor(s_t < \underbar{z})) \bigcup\mathbb{R}^{|\mathcal{X}|-1})$}{
            check criteria compliance by Eq.\ref{eq: verified suffice}
        }
    }
\end{algorithm}

We illustrated a simplified decision tree in Fig.\ref{fig: overview}, where the three decision nodes provide an instance of the box shrunk by the decision comparison rules. If a fail case is detected, we correct it by editing the setpoint in the failed leaf node to the median of the comfort zone. Although this disregards the potential under/overshooting issue, it guarantees that the HVAC system corrects the zone temperature towards the correct direction, as specified in Section \ref{sec: Verification Criteria For HVAC Control}.

\subsubsection{Verifying Criterion \#1.}


This criterion estimates the probability of failure within $H$ time steps starting from a safe state, defined by the forward reachability tube in Eq.~\ref{eq: forward reachable tube 1} given the policy. Again, we utilize the augmented input distribution $\widehat{p(x)}$ that we developed in Section \ref{Sec: Decision Dataset Generation} to sample the more frequent scenarios. One possible procedure to estimate the failure probability is to sample $\widehat{p(x)}$ and run bootstrap predictions to obtain a trajectory of $H$ time steps, then check each step in the trajectory for failure. However, the bootstrap procedure prohibits parallelism and has low computational and space efficiency, since it makes $H$ predictions for each time step in a trajectory only to verify one input. Instead of bootstrapping for $H$ time steps, we show that verifying only one time step ahead is equivalent to the first method with any $H$, and has higher computational efficiency.
\vspace{-7pt}
\begin{proof}
Let $S$ denote the entire input space subject to the \#1 criterion. We divide it into two subsets by the true verification results of its elements, and let $F$ be the set of failed inputs and $N$ be $S\setminus F$ defined by Eq.\ref{eq: verified suffice}. The true verification result of the \#1 criterion is, thusly, $p = \frac{|F|}{|N|}$. Let $R^+(x)|_T$ be the forward reachability tube of $x$ defined in Eq.\ref{eq: forward reachable tube 1}, which contains a trajectory $\{x, \cdots, x_H\}$. We use bootstrap and $x\in N\iff \{x', \cdots, x_H\}\in S$.

Now, instead of bootstrapping, we repeatedly sample start state $x\sim S$ and check if $x' = \hat{f}(x, T(x)) \in S$, such that $x'\in S \iff x \in N$. For any $x\in F$, there will be two cases: $x' \in S\lor x'\not\in S$. If $x'\not\in S$, then $x$ is correctly classified to $F$. Otherwise, there must be another $x_j\in \{x', \cdots, x_H\}$ that is not safe. In this case, $x_{j-1}$ will be classified to $F$, because the immediate next state will be $x_j$, which is not safe. Since $x$ will be classified to $N$ and $x_{j-1}$ to $F$, $|F|$ does not change. Therefore, it correctly estimates the true $\frac{|F|}{|N|}$.
\end{proof}
\vspace{-7pt}
The procedure in the above proof allows more parallelism and fewer model predictions per input. With higher computational efficiency, we verify the first criterion using this method.

\section{Evaluation}



We assess our approach with a high-fidelity simulator in an environment including weather and layout for a fair evaluation.

\subsection{Platform and Implementation Details}\label{sec: Platform and Implementation Details}

\textbf{Softwares.} We used EnergyPlus~\cite{doe2010energyplus} for industrial-level building simulation, PyTorch 2.0.0 for deep learning tasks, Python 3.9, scikit-learn 1.3.2 for decision tree modeling, and Sinergym~\cite{jimenez2021sinergym} for virtual testbed that facilitates interaction with EnergyPlus in Python. Sinergym sends the selected setpoint to the EnergyPlus simulation session, which returns the states back through Sinergym. All software used for our experiment is open source. We used Intel i9-11900KF and GeForce RTX 3080Ti graphic cards for computing.

\textbf{Implementation details.} We used consistent experiment hyperparameters throughout the experiment. For deep learning, we employed settings of epochs=150, learning\_rate = 1e-3, and weight\_decay = 1e-5. We used MSE (Mean Squared Error) as the loss criterion and Adam as the optimizer for all training. For decision data generation, we used noise\_level=0.01. For decision tree construction using CART \cite{loh2011classification}, we left the depth unbounded, and the split threshold was set to its default value. When employing the RS stochastic optimizer, we adopted the optimal hyperparameter configuration as validated in \cite{ding2020mb2c}, specifically sample\_number=1000 and horizon=20.

\textbf{Environment selection.} We conducted our simulation with a $463 m^2$ building with five different zones \cite{jimenez2021sinergym} in two climate-distinct cities from January 1st to January 31st. To ensure generalizability, we selected two cities with distinct climates: Pittsburgh (ASHRAE 4A) and Tucson (ASHRAE 2B), each serving as representatives of unique climate types \cite{standard2020ansi}. For the simulation, we utilized actual 2021 TMY3 weather data specific to these cities \cite{jimenez2021sinergym}.


\subsection{Results}

\subsubsection{Building Control. }

We deploy the policies into the simulated building, monitor the simulation states, and evaluate their performance by energy consumption and violation rate. We select three benchmarks: the building's default rule-based controller~\cite{jimenez2021sinergym}, the MBRL agent \cite{ding2020mb2c}, and the current state-of-the-art method \textit{CLUE} \cite{an2023clue}.

\begin{figure}
\includegraphics[width=\columnwidth]{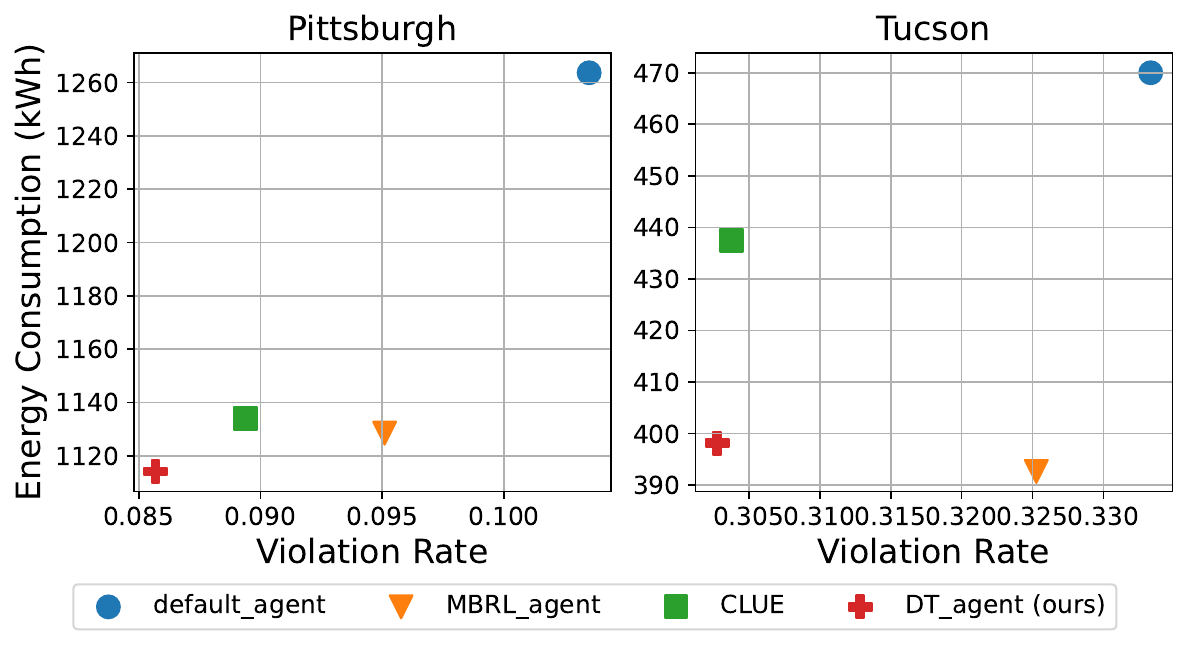}
\caption{Building control results.}
\label{fig: pareto_graph}
\Description[]{}
\end{figure}

We fitted a DT policy for each of the cities and verified them against the proposed criteria. The results are shown in Table \ref{tab: Verification results for two cities.}. Then, we deploy the policies in the simulated buildings and record their building control performances, shown in Fig.~\ref{fig: pareto_graph}. The lower-left direction in Fig.~\ref{fig: pareto_graph} represents the direction of improvement. Compared with the default controller, \textit{CLUE} \cite{an2023clue} saves 129.6 kWh and 32.5 kWh per month for two cities respectively, while our method saves 149.6 kWh and 71.8 kWh, seeing a 68.4$\%$ increase in energy savings and a 14.8$\%$ increase in human comfort on average.

\begin{table}[]
\small
\begin{tabular}{|r|c|c|}
\hline
                                        & Pittsburgh & Tucson \\ \hline
Total No. of nodes                      & 1199       & 3291   \\ \hline
No. of leaf nodes (unique path)         & 599        & 1646   \\ \hline
Safe probability estimated by crit. \#1 & 94.6\%     & 95.1\% \\ \hline
No. of nodes corrected by crit. \#2     & 0          & 0      \\ \hline
No. of nodes corrected by crit. \#3     & 0          & 88     \\ \hline
\end{tabular}
\caption{Verification results for two cities.}
\label{tab: Verification results for two cities.}
\end{table}

\begin{figure}
\includegraphics[width=\columnwidth]{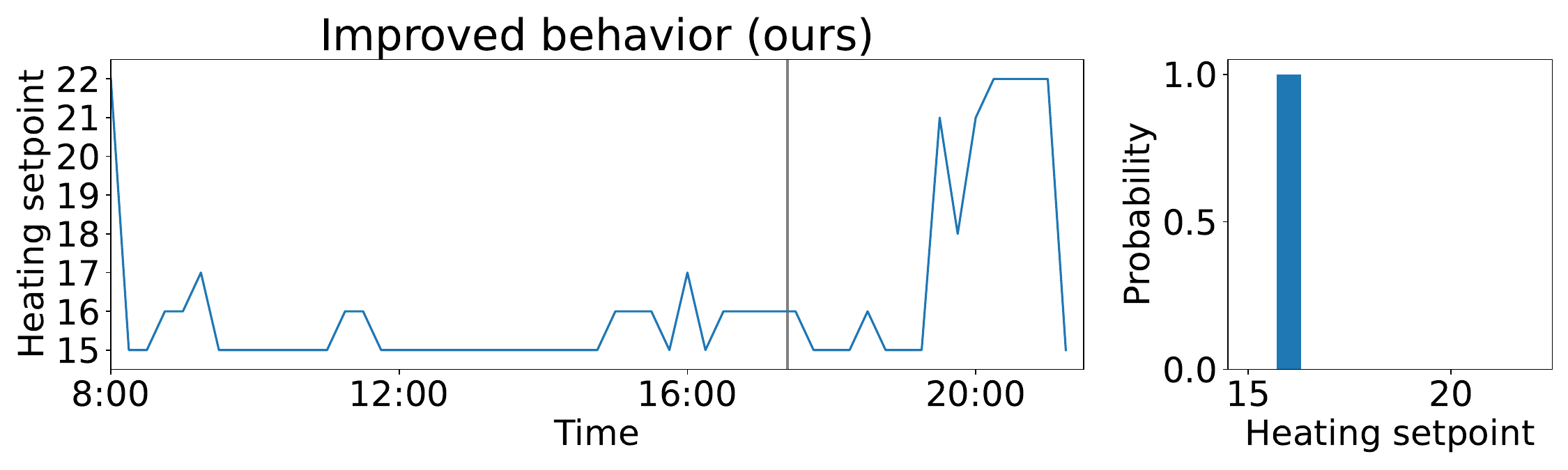}
\caption{Our method's behavior example.}
\label{fig: Jan1_pittsburgh_ours}
\Description[]{}
\end{figure}



Compared with the existing controller, we demonstrate our method's deterministic behavior in the same environment (Fig.\ref{fig: Jan1_pittsburgh_ours}). Despite using MBRL's decisions, our approach is energy-efficient, addressing MBRL's stochasticity by selecting the most frequent action (Section \ref{Sec: Decision Dataset Generation}), confirming its effectiveness in building control.


\subsubsection{Data Efficiency. }

\begin{figure}
\includegraphics[width=\columnwidth]{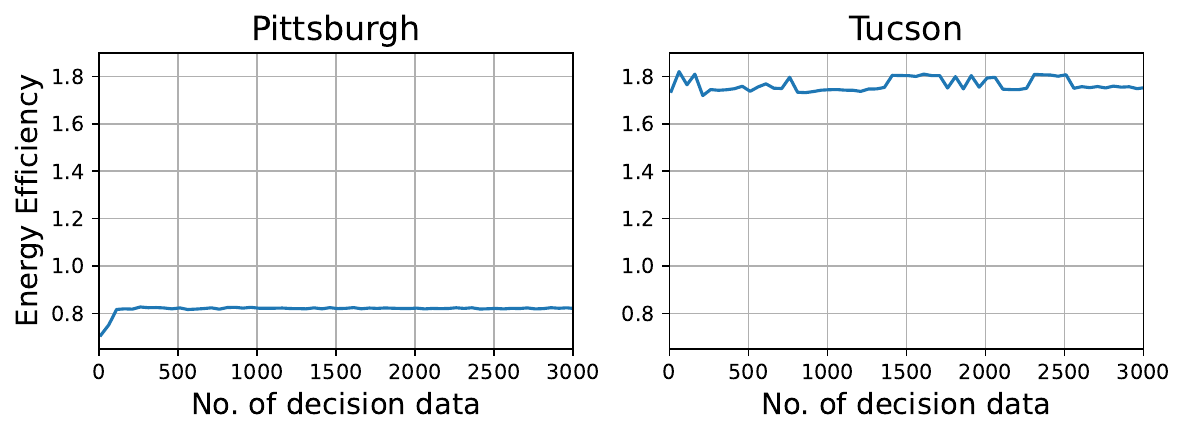}
\caption{Data efficiency results.}
\label{fig: data efficiency}
\Description[]{}
\end{figure}
\begin{figure}
\includegraphics[width=\columnwidth]{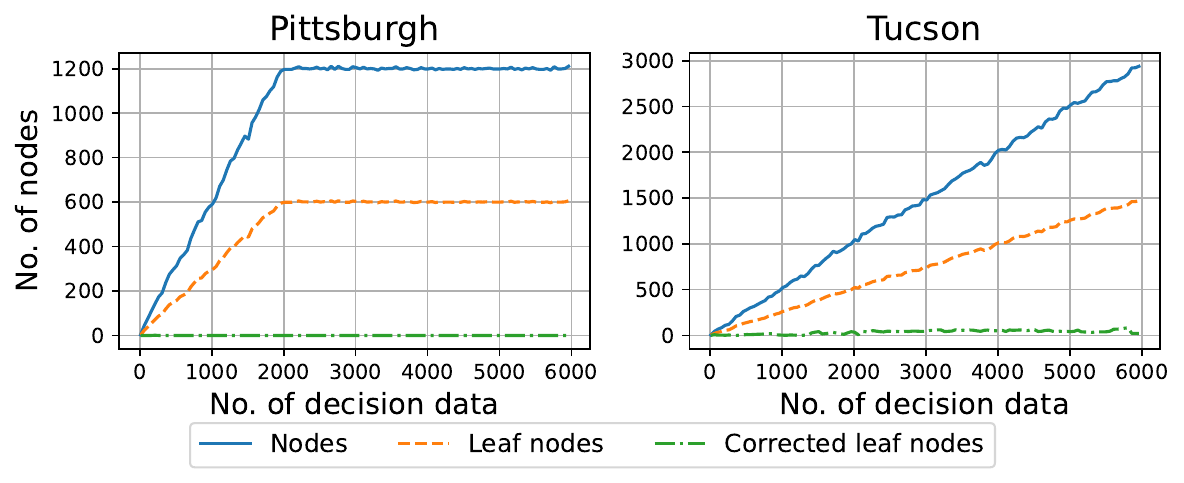}
\caption{Size of DT vs. No. of decision data.}
\label{fig: data efficiency size of tree}
\Description[]{}
\end{figure}

Our proposed DT policy requires an offline decision data generation procedure described in Section \ref{Sec: Decision Dataset Generation} before deployment. In this section, we empirically test the amount of decision data required for our controller to reach optimal performance. We iterate through different numbers of decision data entries, fit a DT policy, deploy the policy to a simulated building, and record its performance. Its performance is measured by the comfort rate divided by the energy consumption, the resulting ratio is multiplied by 1000 for easier presentation. The results are shown in Fig.\ref{fig: data efficiency} and the corresponding sizes of DT are shown in Fig.\ref{fig: data efficiency size of tree}. Fig.\ref{fig: data efficiency} shows that our method converged within 100 decision data points for both cities. Comparing Fig.\ref{fig: data efficiency} and Fig.\ref{fig: data efficiency size of tree} reveals that the DT sizes of both cities converge much later than their building control performances if it converges. This indicates that there is no definitive relationship between DT sizes and control performance. In terms of overhead, the computation time to generate each decision data point averages to 16.8 seconds. This indicates that 28 minutes will be enough to generate sufficient decision data for an optimal DT policy.

\subsubsection{Computation Overhead. }

\begin{table}[]
\begin{tabular}{|r|c|c|c|c|}
\hline
             & default~\cite{jimenez2021sinergym} & MBRL~\cite{ding2020mb2c}   & CLUE~\cite{an2023clue}   & DT (ours) \\ \hline
average (ms) & 0.0            & 212.87 & 326.30 & 0.1888    \\ \hline
std (ms)     & 0.0            & 266.89 & 102.30 & 0.4423   \\ \hline
\end{tabular}
\caption{Online computation overhead results.}
\vspace{-10pt}
\label{tab: Online computation overhead results.}
\end{table}

We recorded the computation overhead of our proposed DT policy and the benchmark methods in the online setting, i.e. assuming that the policy is deployed to a real building. For every method, we record the computation time of each setpoint selection. For our method, we used the model size listed in Table \ref{tab: Verification results for two cities.}. We plot and compare the distribution of the recorded times in Table \ref{tab: Online computation overhead results.}. On average, our method is $1728\times$ and $1127\times$ faster than the previous state-of-the-art methods. This significantly reduces the computation burden of the building edge devices, allowing our method to be deployed to a wider range of buildings.
\section{Conclusion}
 This paper addresses the critical issue of reliability in HVAC systems by introducing innovative approaches rooted in MBRL. By transitioning from black-box models to interpretable decision tree policies and employing domain knowledge-based verification criteria, the research significantly enhances energy efficiency and occupant safety. The utilization of historical data distributions and an offline verification algorithm further solidify the reliability of the control policies. Notably, our method outperforms existing MBRL techniques, offering substantial energy savings, improved comfort, and reduced computational overhead. This work represents a pivotal step toward practical MBRL deployment in the HVAC industry.

 \textbf{Acknowledgements.} This work was supported in part by NSF Grant \#2239458 and UC National Laboratory Fees Research Program Grant \#69763. Any opinions, findings, and conclusions expressed in this material are those of the authors and do not necessarily reflect the views of the funding agencies.



\bibliographystyle{ACM-Reference-Format}
\bibliography{reference}

\end{document}